\documentclass[preprint,aps]{revtex4}

\usepackage{amsmath}
\usepackage{graphicx}
\usepackage{epsfig}
\usepackage{psfrag}
\usepackage{mathdots}
\newlength{\upit}\upit=0.1truein

\newcommand{\ltappr}{{{\lower4pthbox{$<$} } \atop \widetilde{ \ \ \ }}}
\newlength{\bxwidth}\bxwidth=1.5 truein

\begin{document}
\newcommand{\dg}{^{\dagger }}
\newcommand{\vk}{\vec k}
\newcommand{\vq}{{\vec{q}}}
\newcommand{\vp}{\bf{p}}
\newcommand{\al}{\alpha}
\newcommand{\be}{\beta}
\newcommand{\si}{\sigma}
\newcommand{\rarrow}{\rightarrow}
\def\fig#1#2{\includegraphics[height=#1]{#2}}
\def\figx#1#2{\includegraphics[width=#1]{#2}}
\newlength{\figwidth}
\figwidth=10cm
\newlength{\shift}
\shift=-0.2cm
\newcommand{\fg}[3]
{
\begin{figure}[ht]

\vspace*{-0cm}
\[
\includegraphics[width=\figwidth]{#1}
\]
\vskip -0.2cm
\caption{\label{#2}
\small#3
}
\end{figure}}
\newcommand{\fga}[3]
{
\begin{figure}[ht]
\vspace*{-0cm}
\[
\hskip -1.2in\includegraphics[width=\figwidth,angle=+90]{#1}
\]
\vskip -0.2cm
\caption{\label{#2}
\small#3
}
\end{figure}}
\newcommand{\fgb}[3]
{
\begin{figure}[b]
\vskip 0.0cm
\begin{equation}\label{}
\includegraphics[width=\figwidth]{#1}
\end{equation}
\vskip -0.2cm
\caption{\label{#2}
\small#3
}
\end{figure}}

\newcommand \bea {\begin{eqnarray} }
\newcommand \eea {\end{eqnarray}}
\newcommand{\bk}{{\bf{k}}}
\newcommand{\bx}{{\bf{x}}}
\newcommand{\pu}{PuCoGa_{5}}
\newcommand{\np}{NpPd_{5}Al_{2}}

\begin{center}

{\large \bf Heavy electrons and the symplectic symmetry of spin.}\\[0.5cm]

Rebecca ~Flint, M.~ Dzero~ and~ P.~ Coleman\\

{\em
Center for Materials Theory,
Rutgers University, Piscataway, NJ~08854, U.S.A. 
} 
\end{center}\vspace{0.5cm}

{\bf The recent discovery of two heavy fermion materials 
$PuCoGa_{5}$\cite{sarrao} and $NpPd_{5}Al_{2}$\cite{aoki} which
transform directly from Curie paramagnets into superconductors,
reveals a new class of superconductor where local moments 
quench directly 
into a superconducting condensate. 
A powerful tool in the description of heavy
fermion metals is the {\sl large $N$} expansion, which 
expands the physics in powers of $1/N$
about a solvable limit where particles carry a large number ($N$) of
spin components.  As it stands, this method is unable to jointly describe
the spin quenching and superconductivity which develop in 
$\pu $ and $\np $.
Here, we solve this problem with a
new class of large $N$ expansion that employs the symplectic symmetry
of spin to  protect the 
odd time-reversal parity of  spin and 
sustain  Cooper pairs as well-defined
singlets.
With this method we show that when a lattice of magnetic ions exchange 
spin with their metallic environment in two distinct symmetry channels, 
they are able to simultaneously  satisfy both channels 
by forming a condensate of composite pairs between 
between local moments and electrons. 
In the tetragonal crystalline environment relevant to 
$PuCoGa_{5}$ and $NpPd_{5}Al_{2}$ the lattice structure 
selects a  natural  pair of 
spin exchange channels, giving rise to the prediction of a unique 
anisotropic paired state with g-wave symmetry. 
This pairing mechanism predicts a large
upturn in the NMR relaxation rate above $T_{c}$,
a strong enhancement of Andreev reflection in tunneling measurements 
and an enhanced superconducting
transition temperature $T_{c}$ in $Pu$ doped $Np_{1-x}Pu_{x}Pd_{5}Al_{2}$.  }

%
%
%
%
%
%

Large $N$ approximations for interacting electron systems
have provided an invaluable tool for understanding 
heavy fermion materials\cite{read,read2,auerbach}
and low dimensional magnetism\cite{marston,arovas}. 
yet to date, they do not encompass heavy fermion superconductivity. 
We are motivated to return to this unsolved problem by 
the discovery of two new singlet heavy electron superconductors 
$PuCoGa_{5}$
and $\np$\cite{sarrao,aoki}, which transform from Curie paramagnets into
superconductors without first developing a Fermi liquid. 
Unlike other heavy electron superconductors where Cooper pairing is thought to
be driven by antiferromagnetic spin fluctuations
\cite{bealmonod,miyake,scalapino,tanaka}, these materials with a higher
transition temperature do not
appear to be close to a magnetic instability. Moreover, 
the  condensation entropy lost on formation of these superconductor is
between a quarter and a third of the unquenched spin entropy $R \ln 2$
of the Curie paramagnet, indicating that the spin-quenching normally associated
with the Kondo effect is an integral part of the development of
superconductivity.

The difficulty in developing a large $N$ description
of heavy fermion superconductivity is that the odd
time-reversal parity of the electron spin,
$\vec{S}\stackrel{\theta }{\longrightarrow} - \vec{S}$, is not
preserved by $SU (N)$ spins.
The inversion of spins under time-reversal protects singlet superconductivity,
guaranteeing that
an electron paired with its time reversed twin is a singlet. 
However, if we extend the theory so that 
the number of spin components exceeds two, 
the resulting $SU (N)$ spin operators do not all invert under time reversal,
so that time-reversed pairs of particles cease to be spin singlets. 
In this paper we show how the time-inversion of spins 
is restored in a new class of symplectic large $N$ expansions 
which link  time-reversal symmetry 
to a symplectic property of the electron spin.
Our approach enables us to model superconductivity on an equal
footing with the Kondo
effect and leads us to propose that the superconductivity in 
$PuCoGa_{5}$ and $\np$ represents a new kind of lattice-coherence
associated with the Kondo effect, with several observable consequences. 

We begin by demonstrating the link between time reversal and
symplectic symmetry. 
Time reversal is an
antiunitary operator  $\theta $ which 
acts on an electron wavefunction $\psi_{\sigma} (x,t)$ as 
$\theta \psi (\bx ,t) = \hat \epsilon\psi^{*} (\bx ,-t)$
where $\psi^{*}$ is the complex conjugate of the wavefunction 
and $\hat\epsilon_{\alpha \beta }={\rm sgn} ({\alpha })\delta_{\alpha
,-\beta }$ is the skew symmetric matrix which
interchanges ``up'' and ``down'' amplitudes.
$\theta $ is normally written as
a product $\theta = \hat \epsilon K $  \cite{sakurai} with 
the operator $K$, which 
conjugates quantum amplitudes, ${K}\psi  = 
 \psi^{*}{K} $.
Now consistency of time reversal requires that it commutes with spin
rotations, 
$U\theta = \theta U$, or
\begin{equation}\label{condition}
U\epsilon K U\dg = \epsilon K ,
\end{equation}
where $U$ is the unitary rotation operator.  But the conjugate of $U\dg $
is its transpose, $(U\dg )^{*}= U^{T} $, so 
$K$ converts $U\dg$ into its transpose $KU\dg =U^{T}K$,
and consistency requires that 
\begin{equation}\label{symp}
U \hat \epsilon U^{T}= \hat  \epsilon.
\end{equation}
This is the symplectic symmetry of spin: a symmetry that
must hold if the spin rotation group is to remain consistent with the
concept of time reversal.
The unusual appearance of the transpose $U^{T}$, rather than a Hermitian
conjugate $U\dg $ reflects the anti-unitary nature of time reversal.
Replacing $U$ by an infinitesimal rotation, $U = 1
+ i\vec{\alpha }\cdot \vec{ S}$, we obtain the inversion of spins
under time reversal $\vec{S} \stackrel{\theta}{\rightarrow} \theta \vec{S}\theta^{-1}= \hat \epsilon\ 
\vec{S}^{T}\hat \epsilon^T  = - \vec{S}$.

At first sight, this entire set of reasoning immediately extends 
to the case of
fields with  any even number 
$N=2k$ of components, 
$ {\psi}_{\alpha }  \equiv  
\left(
\begin{matrix}
\psi_{1} ~\psi_{-1}
\hdots ~\psi_{k}
~\psi_{-k}
\end{matrix}
 \right)$.  Unfortunately, for $N>2$, the $SU (N)$ group does not satisfy condition (\ref{symp})
for any choice of the matrix $\epsilon$.  To preserve a
consistent definition of both time reversal and spin rotation for $N>2
$, we have to chose the symplectic subgroup $SP (N)$, whose elements
are actually defined by the consistency condition (\ref{symp} ) with
$\hat\epsilon_{\alpha \beta }={\rm sgn} ({\alpha })\delta_{\alpha
,-\beta }$.  Fifteen years ago, Read and Sachdev\cite{readsachdev91}
made the observation that the symplectic group $SP(N)$
allows spin operators that form singlet pairs.  Their approach has
been extensively applied to frustrated
magnetism\cite{oleg,sachdevkagome}, the t-J model \cite{vojta} and
most recently, to paired Fermi
gases\cite{readsachdev91,nikolic,veillette}.  From this discussion,
symplectic spins assume an additional importance as the only 
consistent way to sustain time-inversion symmetry in the large $N$
limit.


The spin operators of the $SU (N)$ group are written
\begin{equation}\label{}
{\cal T}_{\alpha \beta } = \psi \dg_{\alpha }\psi_{\beta }- \left (
\frac{n_{\psi }}{N}\right)\delta_{\alpha \beta },
\end{equation}
where
$n_{\psi }=\sum_{\alpha}\psi\dg_\alpha\psi_\alpha$ is the number of particles that make up the spin. 
Under time reversal,  ${\cal  T}_{\alpha \beta }\stackrel{\theta
}{\rightarrow} \tilde{\alpha }\tilde{\beta} {\cal T}_{-\beta, -\alpha
}$,  $SU (N)$ spins have no well defined parity. 
When $\psi_{\alpha }$ is a Fermi operator, we can also define
a charge conjugation operator $\cal C$ that converts particles into
holes
$\psi_{\alpha
}\stackrel{{\cal C} }{\rightarrow }\tilde{\alpha } \psi \dg_{-\alpha
}$.  
By taking antisymmetric or symmetric combinations of the  $SU (N)$ spins
with their time-reversed version, we may divide them into two
sets 
\begin{eqnarray}\label{sympl}
\hbox{``magnetic'' moments }\qquad S_{\alpha \beta }= \psi \dg_{\alpha}\psi_{\beta }-\tilde{\alpha
}\tilde{\beta }\psi \dg_{-\beta }\psi_{-\alpha }, \qquad 
(\theta ,{\cal C}) = (-,+)
\end{eqnarray}
which invert under time reversal but are neutral under charge conjugation,
and 
\begin{eqnarray}\label{l}
\hbox{``electric'' dipoles}
\qquad \qquad {\cal P}_{\alpha \beta }= \psi \dg_{\alpha}\psi_{\beta }+\tilde{\alpha}\tilde{\beta }\psi \dg_{-\beta }\psi_{-\alpha  }, \qquad (\theta ,{\cal C}) = (+,-)
\end{eqnarray}
which are invariant under time reversal, but
change sign under charge conjugation\cite{zhang}. 
There are $D = \frac{1}{2}N
(N+1)$ magnetic moments that form the generators of $SP (N)$.
For the physical case of $N=2$, 
there are no dipoles, but 
as $N$ becomes large, the $SU (N)$ group contains 
approximately equal numbers of  ``moments'' and ``dipoles''

If we
are to sustain the time-reversal properties of spin in a large $N$
expansion, we must use Hamiltonians that do not contain the unwanted dipole
moment operators. This requirement delineates the new approach from
earlier $SP (N)$ treatments of spin systems.
Since the magnetic moments form a closed
$SP (N)$ algebra, 
their spin dynamics $\frac{dS}{dt}=-i[H[S],S]$  will remain closed and 
decoupled from the magnetic dipoles  provided the Hamiltonian $H$ contains no
dipole moments $\cal  P$ that spoil the closure. 
When the local
moments are built out of fermions, this closure
gives rise to a particularly important gauge symmetry. 
$SU (N)$ spins commute
with the number operator $n_{j\psi }$ at each site, 
giving rise to a $U (1)$ gauge invariance
that features heavily in many  analyses of correlated electron
physics. However, symplectic spins also commute with the fermion pair operator
$\Psi_{j}  = \sum_{\alpha
}\tilde{\alpha}\psi_{j,-\alpha } \psi_{j\alpha }
$ since it is an $SP (N)$ singlet, 
\begin{equation}\label{}
[S_{\alpha \beta },\Psi_{j}
]=[S_{\alpha \beta },\Psi\dg_{j}  ]=[S_{\alpha \beta },n_{j\psi }]=0.
\end{equation}     
In a lattice,  these symmetries
apply  independently at every spin site $j$, giving rise
to an {\sl   $SU (2)$ local  gauge invariance}. 
This 
symmetry was first identified for spin-1/2  by Affleck et 
al.\cite{affleck}, who argued for its 
central role in defining the neutrality of spin.
Symplectic spins allow us to extend this gauge
symmetry to large $N$, provided we exclude 
dipole moment operators from the Hamiltonian. 
This is more stringent requirement than enforcing global $SP (N)$ symmetry, and
to delineate it from earlier $SP (N)$ approaches, we refer
to it as ``Symplectic-$N$''.

A key distinction between earlier $SP (N)$
approaches and Symplectic-$N$, lies in the way spin interactions are
decoupled. The ``dot product''  of symplectic spins $\vec{S}_{i}\cdot \vec{S}_{j}=
\frac{1}{2}S_{\alpha \beta } (i)S_{\beta \alpha } (j)$ 
has a unique decoupling in terms of both particle-hole and singlet 
pairs:
\begin{equation}\label{spinham}
\vec{S}_{1}\cdot\vec{S}_{2} =
- B\dg_{21}B_{21}+ \eta_{\psi } A\dg_{21}A_{21},\qquad  (\eta_{\psi}= \pm),
\end{equation}
where  
$B\dg _{21} 
=\sum_{\sigma }\tilde{\sigma } \psi \dg_{2\sigma }\psi \dg _{1-\sigma }
$
creates a valence bond of spins between sites one and two, while
$A_{21}=
\sum_{\sigma }\psi \dg_{2\sigma }\psi _{1\sigma }$
``resonates''  valence bonds between sites. The signature $\eta_{\psi} $
depends on whether the field 
$\psi $ is a boson ($\eta_{\psi }=+1$) or fermion
($\eta_{\psi }=-1$). 
In the large $N$ limit, these bond variables acquire expectation
values. 
For $N=2$, this kind of decoupling\cite{ceccatto}
is one of many alternative schemes considered by earlier authors.
The standard $SP (N)$ decoupling procedure omits the last
term and is equivalent to a Hamiltonian containing a mixture of spins
and dipoles, $H = J\sum_{i,j} ({\cal S}_{i}\cdot{\cal S}_{j}- {\cal
P}_{i}\cdot{\cal P}_{j})$. 

\vskip 0.1in

As an initial examination of the Symplectic-$N$ procedure 
we compared its performance\cite{supplementary} when applied 
to the frustrated
$J_{1}-J_{2}$ Heisenberg model on a square lattice
\cite{chandradoucot}
\begin{equation}
H = J_{1}\sum_{\bx , \mu}  \vec{S}_\bx  \cdot \vec{S}_{\bx +\mu}+ 
J_{2}\sum_{\bx ,\mu'}\quad \vec{S}_{\bx }\cdot \vec{S}_{\bx +\mu'},
\end{equation}
where $J_{1}$ and $J_{2}$ are the first and next nearest neighbor
couplings.   We used a Schwinger boson representation of the
symplectic spins and compared our results with spin wave theory and
the $SP (N)$ approach.
In classical spin wave theory\cite{chandradoucot,ccl}, the critical spin
$S_{c}$ at which quantum fluctuations melt the Neel order in this lattice
is strongly dependent on the ratio $J_{2}/J_{1}$, diverging at
$J_{2}/J_{1}=1/2$.  In $SP (N)$ theory, where the Heisenberg terms are
expanded purely in terms of pairing fields, the effects of frustration
on the critical spin are not felt, and the critical spin is 
independent of $J_{2}/J_{1}$, an effect presumeably due to the
presence of dipole spin terms with a ferromagnetic
interaction, which over-stabilize the antiferromagnet.
We find that the inclusion 
of valence bond terms in Symplectic-N 
restores the strong dependence of the critical spin on frustration.
Further discussion of this point is found in the online material \cite{supplementary}.

We now turn to the 
application of Symplectic-$N$ to heavy fermion superconductivity in 
$\pu$ and $\np$\cite{sarrao,aoki}, where the appearance
of an $SU (2)$ gauge symmetry  in the fermionic symplectic-$N$
approach has marked physical consequences. 
These materials 
contain a lattice of local moments, immersed in a sea of electrons.
We assume that at low temperatures, the $Pu$ and $Np $ ions in these materials
behave as Kramer's doublets immersed in an electron sea to form a
``Kondo lattice''. 
The exchange of spin with its environment involves 
virtual valence fluctuations  into ionic  configurations
with one more, or one less f-electron:  $f^{n}\rightleftharpoons
f^{n\pm1}\mp e^{-}$, where $n=3$ and $5$ for $\np $ and $\pu$
respectively.  In  complex magnetic ions, 
the partial-wave symmetry of these two virtual processes
are distinct \cite{Cox1996} and we argue that these two processes 
giving rise to two independent Kondo screening channels, 
$\Gamma = 1,2 $, leading to the following model 
\begin{equation}\label{kondo}
\hat{H}=\sum_{{\bk }\sigma}\epsilon_{\bk}c_{\bk \sigma}\dg
c_{\bk \sigma}+
{\textstyle \frac{1}{N}}
\sum_{j}\left[ 
J_1\psi\dg _{{1} \bk \alpha }
\psi_{\Gamma_{1} \bk' \beta }+
J_2\psi\dg _{{2} \bk \alpha }
\psi_{{2} \bk '\beta }
\right] S_{\beta \alpha } (j) e^{i (\bk '-\bk )\cdot {\bf R}_{j}}.
\end{equation}
To develop 
a solvable mean field theory, we examine the family of models  where
\begin{equation}\label{}
\hat S_{\alpha \beta } (j) = f\dg_{j\alpha}f_{j\beta}- \tilde{\alpha
}\tilde{\beta }f\dg _{j\ -\beta }f_{j\ -\alpha}.
\end{equation}
are the $N$ component symplectic representation of 
the magnetic  moment at each site $j$. The physical system corresponds
to the limit $N=2$. 
The quantities $J_{1}$ and $J_{2}$ describe two {\sl unequal} antiferromagnetic
exchange couplings in the two symmetry channels while
the operators 
$\psi\dg _{\Gamma \bk  \alpha }  = \left[ 
\Phi_{\Gamma} (\bk )\right]_{\alpha\sigma}c\dg_{\bk\sigma}$
create electrons  in a partial wave state 
with symmetry $\Gamma = 1,2$,  where the matrices  $\Phi_{1\bk }$ and 
$\Phi_{2\bk }$ are set by the crystal field symmetry.  

In the tetragonal crystal symmetry environment of $\np $
and $\pu $, the  magnetic moments are surrounded by an approximately 
cubic cage of $Pd$ and
$Ga$ ligand atoms, respectively. Electrons that interact with the magnetic
moment must scatter in one of three tetragonal crystal field doublets, 
labelled $\{\Gamma_{7}^{+}, \Gamma_{7}^{-}, \Gamma_{6} \}$. 
The $\Gamma_{7}^{{\pm}}$ states connect selectively with the inplane
and out-of-plane ligand atoms, as shown in Fig. 1.  The $\Gamma_{6}$
crystal field state is aligned along the c-axis, overlapping weakly
with the nearby ligand ions. This leads us to propose a model in which
the two $\Gamma_{7}^{\pm}$ channels dominate the spin exchange with the
f-electrons. 

The presence of two distinct 
scattering channels plays  a central role in our model.   
For a single magnetic Kondo ion, the strongest spin-screening channel
always dominates,  forming a local Fermi liquid of the corresponding symmetry.
At $J_{1}=J_{2}$ the two channels are perfectly balanced,
giving rise to a critical state in which the spin
screening fluctuates between the channels.  Many groups have
speculated that in a lattice environment, the spins will attempt to 
avoid this critical state through the development of 
superconductivity\cite{Cox1996,jarrell,fred,CATK}.
Symplectic $N$ enables us to develop the first controlled realization
of this conjecture, which we apply to the new
superconductors $\pu$ and $\np$.

When we expand the Kondo interaction in (\ref{kondo}), we obtain
$H_{I}=  \sum_{\Gamma= 1,2}H_{\Gamma} (j)$, where
\begin{equation}\label{}
H_{\Gamma} (j)=-\frac{J_{\Gamma}}{N}\left[(\psi\dg _{j\Gamma}f_{j}) 
(f\dg_j\psi_{j\Gamma})+(\psi \dg_{j\Gamma} \epsilon^{\dg }f\dg _j)(f_{j} 
\epsilon \psi _{j\Gamma})
\right]
\end{equation}
describes the spin exchange at site $j$ in channels $\Gamma=1,2$ and
$\psi \dg_{j \Gamma}= \sum_{\bk }\psi\dg_{\bk}e^{-i \bk \cdot {\bf R}_{j}} $ creates an electron in a Wannier state of symmetry $\Gamma$
at site j.
This interaction exhibits the local $SU (2)$ gauge symmetry
$f_{\sigma}\rightarrow
\cos \theta f_{\sigma} +
\sin \theta
 \tilde{ \sigma }f\dg _{\ -\sigma }
$. The important point here, is that this symmetry survives for {\sl all}
even $N$. Earlier efforts have been made to develop $SU (2)$ gauge theories
of heavy electron systems\cite{CATK} and high temperature
superconductors \cite{Wen},  but were not justified
in terms a controlled expansion.

When the Kondo interaction is factorized, it 
decouples into a Kondo hybridization $V$ and pairing field $\Delta$ as follows
\bea
H_{\Gamma} (j)
\to
\sum_{\sigma}
\biggl[\left(f\dg _{\sigma } V_{\Gamma}
+ \tilde{\sigma}f_{\ -\sigma }\Delta_{\Gamma}\right)
\psi _{\Gamma\sigma } +\hbox{H.C}
 \biggr]
+N\left (\frac{
|V_{\Gamma}|^2+|\Delta_{\Gamma}|^2
}{J_{\Gamma}} \right)
\label{decouple}
\eea
where we have suppressed the site indices $j$. The mean-field
Hamiltonian defined by this decoupling becomes exact in the large $N$ limit.
Despite the appearance of ``pairing terms'' in the Kondo interaction, 
the formation of Kondo singlets in a single channel does not lead to
superconductivity; 
an $SU (2)$
gauge  transformation  on the f-electron
can always absorb the pairing term $\Delta $ into a redefinition of
the f-electron:
$
( f\dg _{\sigma } V_{\Gamma}+\tilde{\sigma}f_{-\sigma }\Delta_{\Gamma})\rightarrow 
V_{0}{\tilde{f}\dg }_{\sigma }$. For
a one-channel Kondo lattice, the Symplectic-$N$ and
$SU (N)$ large $N$ limits are thus identical.

For two channels, this is no longer the case. 
Here, it is convenient to
define a Nambu spinor for the f-electrons and two corresponding matrix
$SU (2)$ order parameters 
\begin{equation}\label{composite}
\tilde{f}_{j\sigma } = \begin{pmatrix}f_{\sigma}\cr
\tilde{\sigma}f\dg_{ -\sigma }\end{pmatrix}_{j}, \qquad 
{\cal V}_{\Gamma j}=  \begin{pmatrix}
V_{\Gamma}&\bar \Delta_{\Gamma}\cr
\Delta_{\Gamma}& - \bar V_{\Gamma}
\end{pmatrix}_{j},
\end{equation}
which  transform identically under an $SU (2)$ gauge transformation $g_{j}$,
$\tilde{f}_{j}\rightarrow g_{j}\tilde{f}_{j}$, ${\cal
V}_{\Gamma j}\rightarrow g_{j}{\cal V}_{\Gamma j}$. 
The product ${\cal
V}_{2j}\dg{\cal V}_{1j}$ forms an $SU (2)$ invariant quantity with off-diagonal
component $\Psi=(V_{1j}\Delta_{2j}-V_{2j}\Delta_{1j})$ : this
quantity preserves the local $SU (2)$ invariance, but it {\sl breaks} the
global $U (1)$ gauge invariance associated with physical charge, and
plays the role of a superconducting order parameter.
If we carry out
an $SU (2)$
gauge transformation that removes the $\Delta_{1j}$, the composite
order parameter $\Psi_{j}= V_{1j}\Delta_{2j}$. These terms lead to Andreev reflection off the screened Kondo
impurity, as shown in Fig. 1b.

Physically, we may understand this phenomenon as the consequence
of the formation of a condensate of composite pairs: 
pairs of electrons, bound to magnetic impurities.  When an electron scatters off a magnetic
impurity, the quantity $\Psi $ determines the amplitude
for emitting an Andreev hole, leaving behind a composite pair.
Detailed analysis\cite{supplementary}
confirms this insight and demonstrates the formation of composite
order with expectation value
\begin{equation}\label{}
\langle \Psi_{N-2}\vert
\psi^{1}_{\downarrow}(j)
\psi^{2}_{\downarrow}(j){S}^+_{f}(j)| \Psi_{N} \rangle.
\propto (V_{1j}\Delta_{2j}-
V_{2j}\Delta_{1j})
\label{orderparameter}.
\end{equation}
In a single impurity model, this order parameter is forbidden,
because electrons can never change between scattering channels, 
but in the lattice, electrons
travelling between sites no longer conserve the channel index, 
which  permits composite order.  Once composite order develops, the
Kondo singlets resonate between the two screening channels, and the
resonance energy this gives rise to stabilizes the coherent state. 

This basic mechanism was 
was first proposed in \cite{CATK}; 
our large $N$ analysis provides the first 
solvable limit in where this mechanism can be
rigorously validated, while at the same time incorporating the
detailed spin-orbit physics of the crystal field-split screening channels.
When we evaluate the resonant Andreev scattering 
from the Feynman diagram
shown in (\ref{fig1}), we find that the conduction electrons acquire a gap
that is proportional to the overlap of the two hybridization functions
$\Delta_{e} (\bk )\propto
{\rm Tr}[\Phi\dg _{2\bk }\Phi_{1\bk }]$, as  shown in Fig. 1. One of
the fascinating features of this gap, is that it contains nodes,
despite the absence of any such nodes in the underlying
hybridization.  This is a unique consequence of the 
the hybridization between $\vert\Gamma_{7}^{+}\rangle 
\sim \vert \pm
3/2\rangle $ and $\vert\Gamma_{7}^{-}\rangle \sim
\vert \mp 5/2\rangle 
$ electron states in the
composite pair. Since these states differ by $m=\pm 4$ units of
angular momentum, this gives rise to  composite pairs 
with $l=4$ units of angular momentum that contains gap nodes.

We have computed the superconducting transition temperature $T_{c}$ as
a function of the ratio $J_{2}/J_{1}$ using Symplectic-$N$\cite{supplementary}. 
Fig. 2 (b) shows the results of a model calculation for a two-dimensional
Kondo lattice, assuming uniform expectation values for ${\cal V}_{\Gamma}$.
It is instructive to
contrast the phase diagrams of the $SU (N)$
and symplectic large $N$ limits.
In the former, there is a single quantum phase
transition that separates the heavy electron Fermi liquids formed via a
Kondo effect about the strongest channel. In the symplectic treatment,
this quantum critical point is  immersed beneath a
superconducting ``dome''. The Cooper channel in the heavy electron normal state
guarantees that the secondary screening channel is always marginally
relevant in the lattice, as first speculated in \cite{CATK}. 
This is, to our knowledge, the first {\sl controlled}
mean-field theory in which the phenomenon of ``avoided criticality''
gives rise to superconductivity in a model that can be solved exactly.

There are three concrete consequences of composite pairing that permit
our ideas to be compared with experiment: 
\begin{enumerate}

\item {\sl Crystal fields determine the gap symmetry.}  
When an electron scatters
between the  $\Gamma = \Gamma_{7}^{\pm}$ channels, it scatters between
the $\vert \pm 3/2\rangle \leftrightarrow \vert \mp
5/2\rangle $ states, and in so doing
picks up $l=4$ units of angular momentum via spin-orbit scattering.
The corresponding
order parameter then acquires the symmetry of an  orbital state with
$l=4$,  and the corresponding order parameter thus has $g$ (l=4)- wave
symmetry, with eight nodal surfaces, as shown in Fig. 1. 
The symmetry of this gap is independent
of the microscopic details, and purely set by the tetragonal
crystalline environment of the magnetic ions. 

\item {\sl Upturn in the NMR.}
In the approach to the composite ordering transition, the local moments
must correlate between sites, and this manifests itself through the
development of an enhancement of the NMR relaxation rate. 
For those systems with maximal $T_{c}$, 
the NMR relaxation rate in the normal state is predicted to contain a
term derived from the interference between the two screening channels, 
proportional to the product  $\frac{1}{T_{1}T}\propto J_{1}
(T)J_{2} (T)$ of the temperature-renormalized Kondo coupling constants.
This gives rise to
an upturn in the NMR relaxation rate $\frac{1}{T_{1}T}\propto [\ln
^{2} (T/T_{c})+ \pi^{2}]^{-1}$, a result in accord with recent
measurements on $\pu $ \cite{Curro2005}, but which has yet to be
tested in $\np $.

\item {\sl Andreev reflection.} The main driver for this mechanism of
heavy fermion superconductivity is Andreev reflection off quenched
magnetic moments. Conventional Andreev reflection
involves the direct transfer of an electron from the probe
into the pair condensate of the conduction sea. 
The conventional BTK theory of  Andreev reflection \cite{btk}
predicts that such processes are severely
suppressed by the large mis-match between the probe and heavy electron
group velocities. 
However, in a heavy electron system, an electron can also ``cotunnel'' into
the Kondo lattice - a process well-known in magnetic quantum dots, whereby
the electron flips a localized spin as it tunnels into the
material\cite{glazman}.  
In a composite paired superconductor, these processes will
result in the direct absorption of the electron into the condensate of
composite pairs, giving rise to an enhanced
contribution to the Andreev tunnel current with a Fano resonant structure.

\item {\sl Internal Proximity effect.} When the Kondo ions in the
superconductor are substituted by Kondo ions with a larger coupling constant,
the robust nature of the gap symmetry will protect the superconductor
against pair-breaking, and is expected 
to lead to  an internal proximity effect,
where the Andreev reflection off the substituted impurities 
{\sl enhances} the superconducting $T_{c}$. This effect requires an
overlap between the gap functions of the two different ions, but this
is guaranteed by crystal symmetry, providing the same screening channels
are operative for both ions.  Based on this line of argument, 
we expect 
that $Pu$ doping of $\np $ will lead to an enhancement of
the superconducting $T_{c}$.

\end{enumerate}

Our work currently leaves open, the question of the
link between $\np $ and $\pu $ and other 
actinide and cerium
systems of similar tetragonal structure, including 
${PuRhGa_{5}}$\cite{purhga5},  $CeRhIn_{5}$\cite{cerhin5}, $CeIrIn_{5}$\cite{ceirin5} and
 $CeCoIn_{5}$\cite{cecoin5}. Each of
these systems develops superconductivity, but magnetic
susceptibility measurements indicate that the f-moments are more completely
quenched at the superconducting transition.  We are tempted to suggest
that that these systems are examples of composite pairing in the
parameter range where $J_{2}/J_{1}$ is smaller, and further away from
the maximum transition temperature. Neutron scattering measurements do 
indeed shown that the two lowest lying crystal field states in
the Cerium $115$ materials are the $\Gamma_{7}^{+}$ and
$\Gamma_{7}^{-}$ states, respectively\cite{xf115}.
Moreover, 
strong Andreev reflections have recently been observed in the tetragonal
$Ce$ 115 heavy electron superconductors, a phenomenon that
could be associated with cotunneling into the composite pair condensate
\cite{greene}.
However, at the same time,
superconductivity in these systems clearly develops in close proximity to
antiferromagnetism, so we can not rule out 
antiferromagnetic spin fluctuations as the predominant pairing
mechanism  in these cases. 
Future work is needed
to examine whether the superconductivity 
observed in these materials is
intimately linked to the tetragonal crystal field structure, as would be
expected in our model.


In summary, by making the observation of a close link between
time-reversal and symplectic symmetry, we have proposed a new class of
large $N$ expansion that permits a controlled treatment of the Kondo
effect and superconductivity on an equal footing.  Using this
approach, we have proposed a model for the heavy
electron materials $\np $ and $\pu $, which attributes the
simultaneous development of superconductivity and spin-quenching  
to the development of a condensate of composite  pairs that give rise
to resonant bound-states that develop a coherent Andreev component
to the electron scattering, with a variety of
experimental consequences.  We note in finishing that the methods used
in this paper may also be applied to one-band electronic
systems\cite{vojta}, an aspect of interest for future work.

This research was supported by the National Science Foundation grant
DMR-0605935 and DE-FE02-00ER45790 (M. Dzero).  The authors would like
to thank N. Andrei, K. Basu, E. Miranda, R. Moessner, P. Pagliuso, S. Sachdev, 
N. Read, G. Zarand and particularly Scott Thomas, for discussions
related to this work.



\figwidth=16cm

\fg{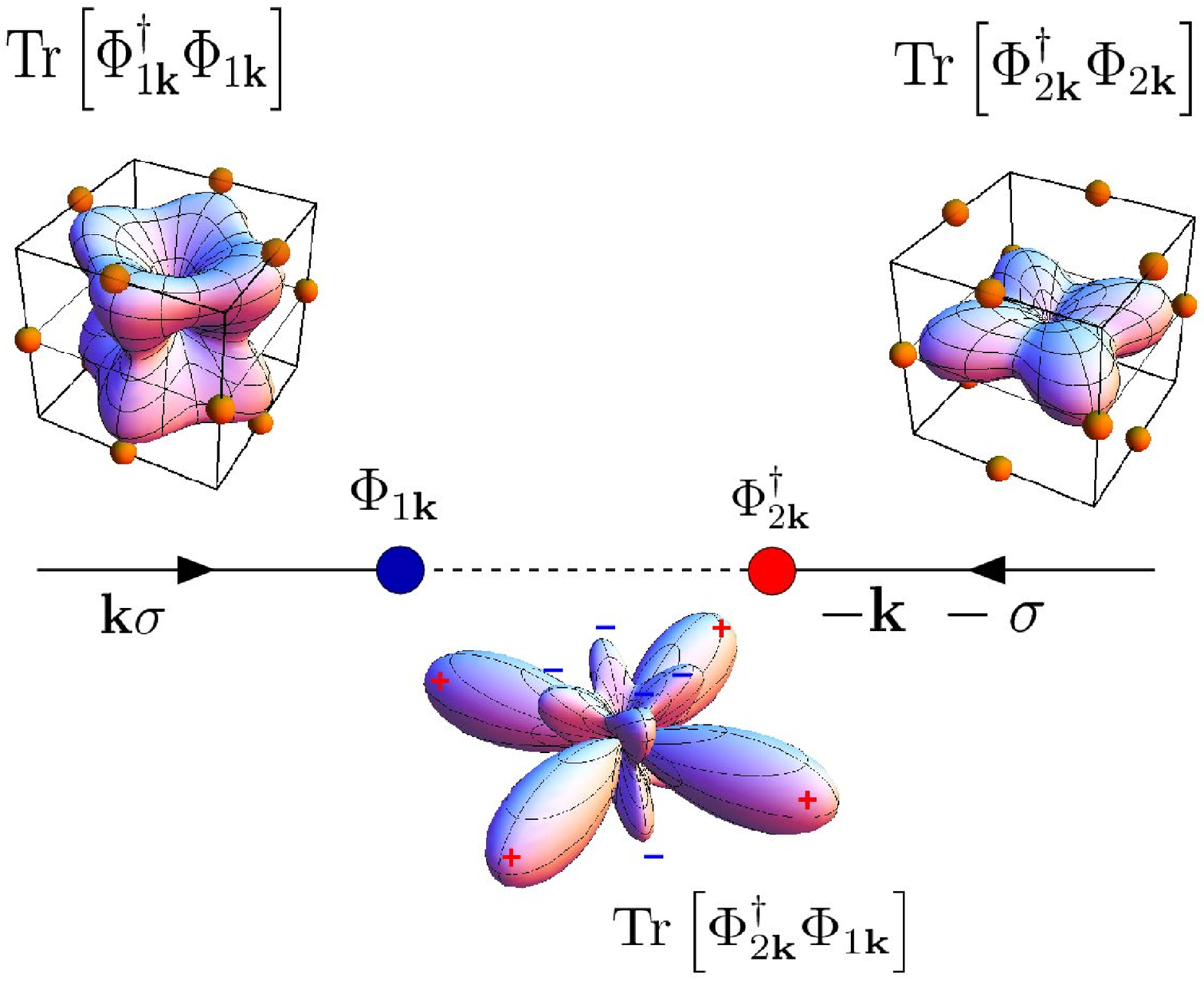}{fig1}{(Top) Showing the angular dependence of 
of the hybridization function
$|\Phi_{\Gamma \bk }|^{2}$ for the two tetragonal crystal field
scattering states,  $\Gamma_{7}^{+}$ and  $\Gamma_{7}^{-}$ with
$j=5/2$. The $\Gamma_{7}^{\pm}$ states are linear combinations of
$\vert \Gamma_{7}^{+}\sigma \rangle = \cos \theta \vert \mp 3/2\rangle +
\sin\theta \vert \pm 5/2\rangle $ 
and 
$\vert \Gamma_{7}^{-}\sigma \rangle = \cos \theta \vert \pm 5/2\rangle -
\sin\theta \vert \mp 3/2\rangle $. Figures display the crystal field
configurations with maximum overlap with the in-plane
and out-of-plane ligand atoms, which  occurs for $\theta \approx
\pi/10$. 
(Centre:) Formation of a composite pair between the two channels leads to
Andreev  scattering of conduction electrons between the two channels,
generating a superconducting gap with symmetry given by the the trace
of the overlap between the two hybridization functions.  (Bottom:) The transfer
of four units of orbital angular momentum that  accompanies this process
leads to a gap with $l=4$, ``g-wave'' symmetry, with four large lobes
of equal sign in the basal plane, as illustrated.
}

\hskip -0.5truein\fga{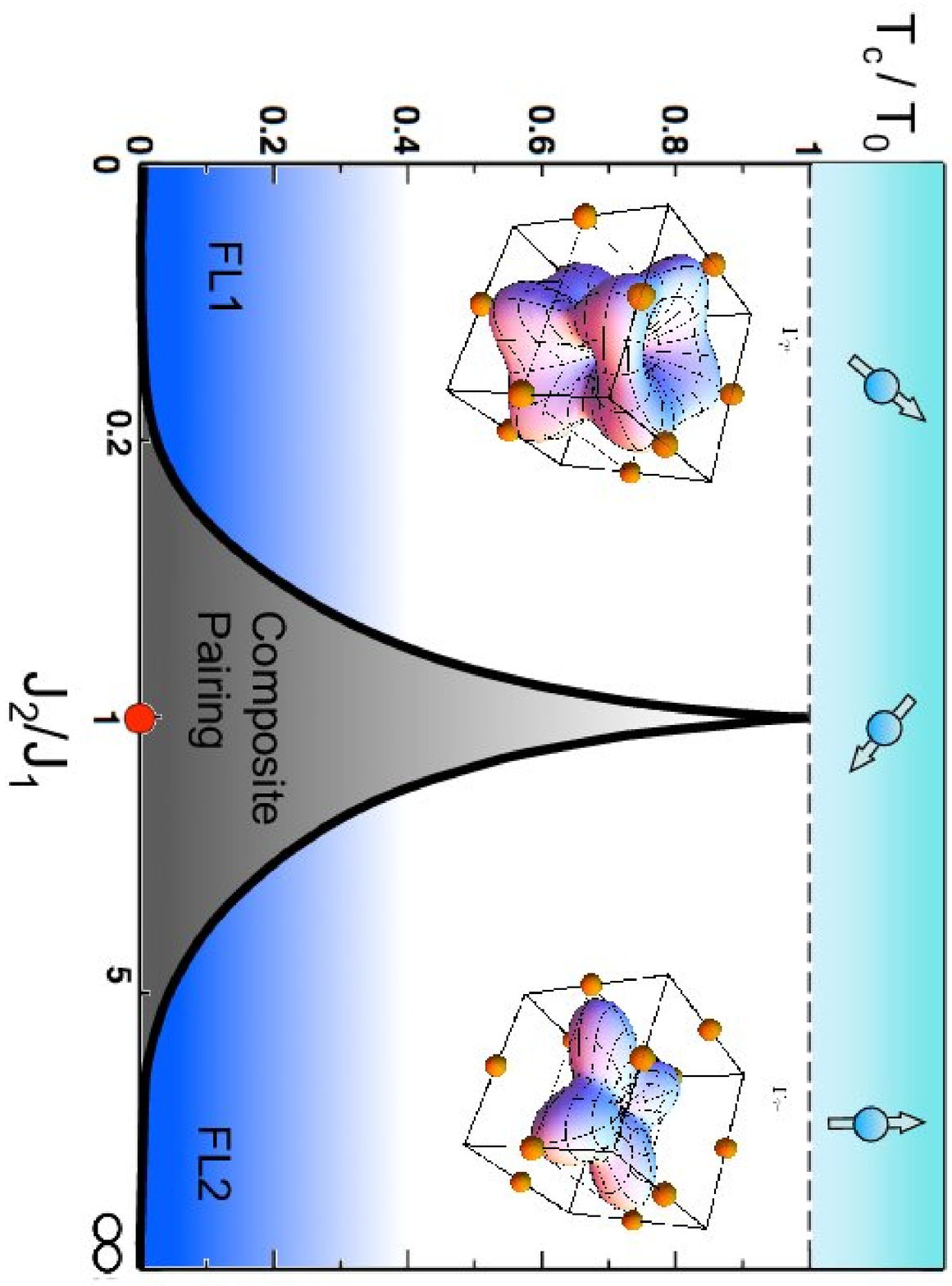}{fig2}{
Phase diagram for two channel
Kondo lattice, computed in the symplectic large $N$ limit for a
tetragonal symmetry  Kondo lattice in which spin is exchanged in
channels $\Gamma_{1}\equiv \Gamma_{7}^{+}$ and
$\Gamma_{2}\equiv \Gamma_{7}^{-}$.
The x-axis co-ordinate is the parametric variable  $x=2
(J_{2}/J_{2})/ (1+J_2/J_1)$ running from $x=0$ to $x=2$, corresponding
to $J_{2}/J_{1}$ running from zero to infinity, as labelled.
Temperature is measured in units of the maximum Kondo temperature of
the two channels $T_{0}= {\rm max} (T_{K1},T_{K2})$. 
Two Fermi liquids of different symmetry develop in the regions of
small, or large $J_{2}/J_{1}$, separated by a common region of
composite-pairing, delineated by the gray area of the phase diagram.
The red-point denotes the location of the single-impurity 
quantum critical point that develops when the two channels are
degenerate.
In the lattice, this point is avoided through the development of
composite pairing. 
}

\newpage
\fg{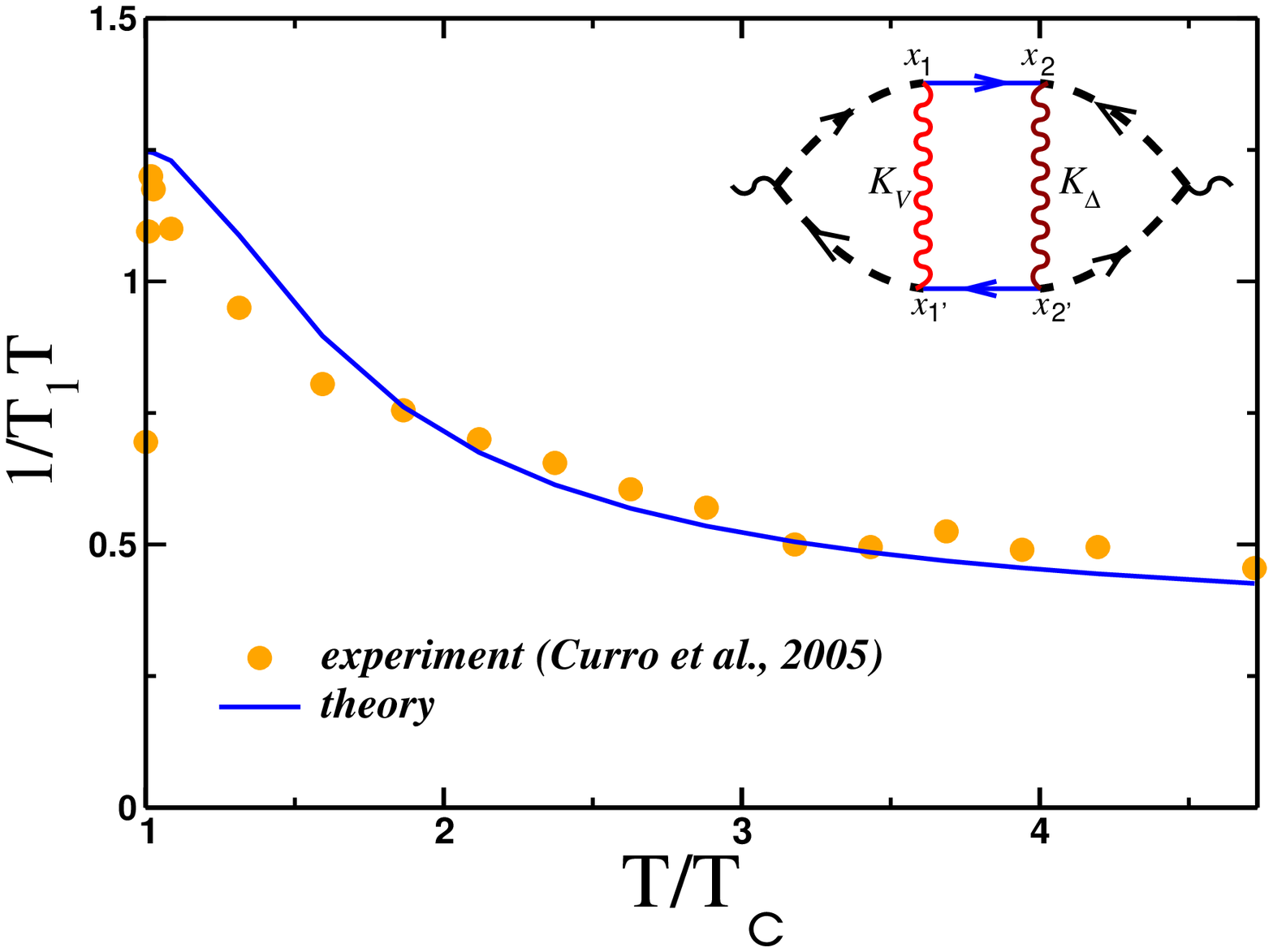}{fig3}{
Upturn in the NMR relaxation rate created by the co-operative 
interference of the Kondo  effect in the two channels at different
sites,  for the extreme case of maximum $T_{c}$, where
$J_{1}=J_{2}$ (Blue line), 
compared with measured NMR relaxation rate in $\pu $ \cite{Curro2005}
(yellow points). Inset shows Feynman diagram used to compute this contribution, 
where dotted lines describe the f-fermions, blue lines describe
conduction electrons propagating between sites and the curly lines
describe the Kondo interaction  in the particle-hole (red) and
particle-particle (blue) channels at different sites.  Temperature is
measured in units of the transition temperature $T_{c}$
\cite{supplementary}
}

\begin{thebibliography}{99}

\bibitem{sarrao} J. L. Sarrao \emph{et al.}, ``Plutonium-based superconductivity with a transition temperature above 18 K", Nature (London) {\bf 420}, 297 (2002). 

\bibitem{aoki}D. Aoki et al, ``Unconventional Heavy-Fermion Superconductivity of a New Transuranium Compound $NpPd_{5} Al_{2}$'',Journal of the Physical Society of Japan, {\bf{76}}, 063701 (2007).

\bibitem{read}N. Read, and D. M. Newns,''A new functional integral formalism for the degenerate Anderson model", {\it J. Phys. C} {\bf 16}, L1055, (1983).

\bibitem{read2} N. Read and D.M. Newns,''On the solution of the Coqblin-Schrieffer Hamiltonian by the large-N expansion technique", {\it J. Phys. C} {\bf 16}, 3273 (1983).

\bibitem{auerbach}A. Auerbach, and K. Levin, 
``Kondo Bosons and the Kondo Lattice: Microscopic Basis for the Heavy Fermi Liquid",
{\it Phys. Rev. Lett.} {\bf  57}, 877 (1986).

\bibitem{marston} I. Affleck and J. B. Marston, ''Large-$N$ 
limit of the Heisenberg-Hubbard model: Implications for high-T$_c$ superconductors'',
{\sl Phys. Rev. B}, {37}, {3774--3777},
{(1988)}.

\bibitem{arovas}
D. P. Arovas and  A. Auerbach, ``Functional integral theories of 
low-dimensional quantum Heisenberg models'',
\newblock {\em Phys. Rev. B}, {\bf 38}, 316, (1988).


\bibitem{bealmonod}
M. T. Beal Monod, C. Bourbonnais, C., and V. Emery, ``Possible
superconductivity in nearly antiferromagnetic itinerant fermion systems'', {\em Phys. Rev. B.} {\bf 34} 7716, (1986). 

\bibitem{miyake}
K. Miyake, S. Schmidt Rink and C. M. Varma, ``Spin-fluctuation-mediated even-parity pairing in heavy-fermion superconductor'', {\em Phys. Rev. B}, 34, 6554,  (1986).

\bibitem{scalapino} D. J. Scalapino, D.~J., E. Loh  and J. E. Hirsch,
``Possible superconductivity in nearly antiferromagnetic itinerant fermion systems'', \newblock {\em Phys. Rev. B}, {\bf34}, 8190, (1986).

\bibitem{tanaka}K. Tanaka, I. Hiroaki and K. Yamada, ``Theory of
superconductivity in $PuCoGa_{5}$'', J. Phys. Soc. Japan {\bf{73}}, 1285 (2004).

\bibitem{sakurai}J. J. Sakurai, Modern Quantum Mechanics Revised Edition, pp 277, (Addison Wesley), (1994).


\bibitem{readsachdev91}N. Read and Subir Sachdev,''Large-N  expansion for frustrated quantum antiferromagnets'',
{\it Phys. Rev. Lett.} {\bf  66}, 1773 (1991); Subir Sachdev and Ziquiang
Wang, ``Pairing in two dimensions: A systematic approach'', {\it Phys. Rev.} B {\bf  43}, 10229 (1991).

\bibitem{oleg}O. Tchernyshyov, R. Moessner and S. L. Sondhi,
``Flux expulsion and greedy bosons: Frustrated magnets at large-$N$'',
{\it Europhysics Letters} 73, 278-284 (2006).

\bibitem{sachdevkagome}S. Sachdev, ``Kagom\'{e}- and triangular-lattice Heisenberg antiferromagnets: Ordering from quantum fluctuations and quantum-disordered ground states with unconfined bosonic spinons'',{\it Phys. Rev. } {\bf  45}, 12377 (1992).


\bibitem{vojta}Matthias Vojta, Ying Zhang, Subir Sachdev, ''Competing
orders and quantum criticality in doped antiferromagnets'',  Phys. Rev. B 62, 6721 (2000).




\bibitem{nikolic} P. Nickoli\'c and S. Sachdev, ``Renormalization-group fixed points, universal phase diagram, and $1/N$ expansion for quantum liquids with interactions near the unitarity limit'', {\it Phys. Rev. }A {\bf 75}, 033608 (2007).

 \bibitem{veillette} M. Veillette, D. Sheehy and L. Radzihovsky, ``Large-$N$ expansion for unitary superfluid Fermi gases'', {\it Phys. Rev.} A {\bf 75}, 043614 (2007).


\bibitem{zhang}C. Wu and S. Zhang, ``Sufficient condition for absence of the sign problem in the fermionic quantum Monte Carlo algorithm'', Phys. Rev. B {\bf 71}, 155115 (2005).

\bibitem{affleck} I. Affleck, Z. Zou, T. Hsu and P. W. Anderson,
``SU(2) gauge symmetry of the large-U limit of the Hubbard model'', {\it Phys. Rev. B} {\bf  38}, 745, (1988).


\bibitem{ceccatto} H.A. Ceccatto, C.J. Gazza and A.E. Trumper, ``Nonclassical disordered phase in the strong quantum limit of frustrated antiferromagnets'', {\it Phys. Rev. B} {\bf   47}, 12329 - 12332 (1993).

\bibitem{supplementary} ``Supplementary Information is linked to the online version of the paper at www.nature.com/nature."

\bibitem{chandradoucot} P. Chandra and B. Doucot, ``Possible spin-liquid state at large-$S$ for the frustrated square Heisenberg lattice'',{\it Phys. Rev. } B{\bf  38}, 9335, 1988

\bibitem{ccl}P. Chandra, P. Coleman and A. I. Larkin, ''Ising transition in frustrated Heisenberg models'', {\it Phys. Rev. Lett.} {\bf 64}, 88 (1990).

\bibitem{Cox1996} D. L. Cox and M. Jarrell, ``The two-channel Kondo route to non-Fermi-liquid metals'', J. Phys.: Condens. Matter {\bf 8}, 9825 (1996).

\bibitem{jarrell}   Mark Jarrell, Hanbin Pang and D. L. Cox, ``Phase Diagram of the Two-Channel Kondo Lattice'', Phys. Rev. Lett, 78, 1996  (1997).

\bibitem{fred}See D. L. Cox and A. Zawadowski, ``Exotic Kondo effects
in metals: magnetic ions in a crystalline electric field and tunneling
centers'', Advances in Physics, 47 599 (1998) and references therein.

\bibitem{CATK} P. Coleman, A. M. Tsvelik, N. Andrei and H. Y. Kee, ``Co-operative Kondo effect in the two-channel Kondo lattice'', {\it Phys. Rev. B }{\bf 60}, 3608 (1999).

\bibitem{Wen} Y. Ran and X.G. Wen, ``Continuous quantum phase transitions beyond Landau's paradigm in a large-$N$ spin model'', cond-mat/0609620v3 (2006).






\bibitem{Curro2005} N. J. Curro \emph{et al.}, ``Unconventional superconductivity in PuCoGa$_5$'', Nature (London) {\bf 434}, 622 (2005).

\bibitem{btk} G. E. Blonder, M. Tinkham and T. M. Klapwijk,"Transition from metallic to tunneling
regimes in superconducting microsconstrictions: Excess current, charge imbalance, and supercurrent conversion", Phys. Rev. {\bf B 25}, 4515 (1982).

\bibitem{glazman} A. Kaminski, Yu. V. Nazarov, and L. I. Glazman,
''Suppression of the Kondo Effect in a Quantum Dot by External Irradiation'', Phys. Rev. Lett. 83, 384 - 387 (1999).

\bibitem{purhga5}F. Wastin, P. Boulet, J. Rebizant, E. Colineau, and
G. H. Lander, ``Advances in the preparation and characterization of
transuranium systems'', J. Phys.: Condens. Matter 15, S2279 (2003).

\bibitem{cerhin5}H. Hegger, C. Petrovic, E. G. Moshopoulou,
M. F. Hundley, J. L. Sarrao, Z. Fisk, and J. D. Thompson,''Pressure-Induced Superconductivity in Quasi-2D $CeRhIn_{5}$'' Phys. Rev. Lett. 84, 4986 (2000).

\bibitem{ceirin5} C. Petrovic, R. Movshovich, M. Jaime, P. G. Pagliuso, M. F. Hundley, J. L. Sarrao, Z. Fisk, and J. D. Thompson,''A new heavy-fermion superconductor CeIrIn5: A relative of the cuprates?'' Europhys. Lett. 53, 354 (2001).

\bibitem{cecoin5}C. Petrovic, P. G. Pagliuso, M. F. Hundley, R. Movshovich,
J. L. Sarrao, J. D. Thompson, Z. Fisk, and P. Monthoux, ``Heavy-fermion superconductivity in CeCoIn5 at 2.3 K'', J. Phys.:
Condens. Matter 13, L337 (2001).

\bibitem{xf115}A. D. Christianson, E. D. Bauer, J. M. Lawrence,
P. S. Riseborough, N. O. Moreno, P. G. Pagliuso, J. L. Sarrao,
J. D. Thompson, E. A. Goremychkin, F. R. Trouw, M. P. Hehlen,
R. J. McQueeney, ``Crystalline Electric Field Effects in CeMIn5:
Superconductivity and the Influence of Kondo Spin Fluctuations'',
Phys. Rev. B 70, 134505 (2004).

\bibitem{greene}W. K. Park, J. L. Sarrao, J. D. Thompson and
L. H. Greene, ``Andreev Reflection in Heavy-Fermion Superconductors
and Order Parameter Symmetry in $CeCoIn_5$'', arXiv.org:0709.1246, (2007).


















\end{thebibliography}
\end{document}